# Complete Two-dimensional Muellermetric Imaging of Biological Tissue Using Heterodyned Optical Coherence Tomography

Xue Liu[1], Shih C. Tseng[1] and M.S. Shahriar[1,2]

[1]*Department of Electrical Engineering and Computer Science, Northwestern University, Evanston, IL 60208*

[2]*Department of Physics and Astronomy, Northwestern University, Evanston, IL 60208*

A polarization-sensitive optical coherence tomography system based on heterodyning and filtering techniques is built to perform Stokesmetric imaging of different layers of depths in a porcine tendon sample. The complete 4×4 backscattering Muellermetric images of one layer are acquired using such a system. The images reveal information indiscernible from a conventional OCT system.

**Introduction**

Optical coherence tomography (OCT) is widely used in biological sub-surface imaging due to its non-contact, non-destructive properties and high resolution [1,2,3,4]. Many biological tissues, such as tendon, bone, and tooth, exhibit birefringence because of their linear or fibrous structure, which alters the polarization state of light propagating in them [5,6]. For the purpose of acquiring the polarimetric signatures of biological tissues, several polarization-sensitive OCT (PSOCT) systems have been developed in recent years [7,8,9,10,11]. For example, de Boer *et al.* used the PSOCT system to generate images of thermally damaged tissue [7]. Hitzenberger *et al.* used the PSOCT to detect the phase retardation and fast axis orientation in chicken myocardium [8]. Everett *et al.* applied the PSOCT to the measurement of the birefringence of porcine myocardium [10].

It is well-known that the Mueller matrix fully characterizes the polarimetric signature of the object [12]. The 16 unique elements of the Muller matrix of different media can be used for object characterization and identification [13,14,15]. Yao *et al.* reported a PSOCT system aiming to yield the full Mueller matrix of the bone of a yellow croaker [16]. However, the PSOCT system used in that experiment is capable of capturing only the polarized light in the sample reflection.

Recently, we demonstrated a PSOCT system capable of measuring the complete Stokes vector of partially polarized light using heterodyning and filtering techniques [17]. In this paper, we apply such a PSOCT system to perform Stokesmetric imaging of different layers in a pork tendon sample. The full 4×4 Muellermetric images of one particular layer in the sample are investigated with the PSOCT system.

**PSOCT system**

Fig.1 depicts the configuration of the heterodyned polarization-sensitive OCT system. A collimated beam from a broadband superluminescent diode (SLD) at a central wavelength of 830nm is used as the light source. The linear polarizer (LP) selects a purely linear input state of the beam before being launched into the acousto-optic modulator (AOM). The unshifted beam at central frequency $\omega_b$ and the first order diffraction at central frequency $\omega_a$ are coupled into the two arms of a modified Michelson interferometer as the incident sample beam and the reference beam, respectively. A half-wave plate (HWP), followed by a quarter-wave plate (QWP) is placed on each arm of the Michelson interferometer to control the polarization state of the light. The probe beam is focused by a lens onto the sample mounted on a two-dimensional motorized translation stage, which can be driven for vertical and horizontal scans by computer programs. The backscattered light from the sample recombines with the reference beam at a non-polarizing beam splitter (BS1). Only when the optical path difference between the two arms is within the coherence length of the SLD source do the two beams produce a beat note at the central frequency $\omega = \omega_a - \omega_b$ [18]. Thus, by moving M2 axially on the translation stage, one can select backscattered light from different depth layers of the sample to heterodyne with the reference beam. A photodetector is placed after BS1. The detected signal is sent into a high pass filter (HPF) followed by a mixer and a low pass filter (LPF). The HPF, used to extract the beat note at frequency $\omega$, has a low frequency cutoff larger than the bandwidth of any low-frequency system noise. The bandwidth of the LPF is set lower than the depolarization bandwidth in order to extract the DC signal after the mixer.

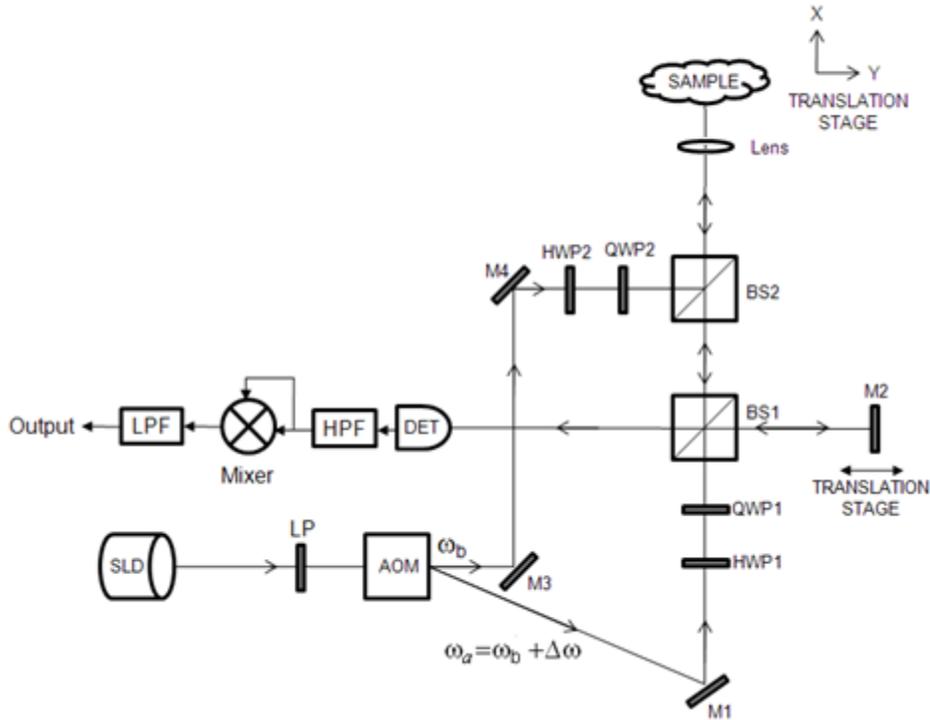

*Fig. 1. Schematic diagram of the PSOCT system. SLD: superluminescent diode; LP: linear polarizer; AOM: acousto-optic modulator; HWP: half wave plate; QWP: quarter wave plate; BS: beam splitter; M: mirror; DET: detector; HPF: high-pass filter; LPF: low-pass filter.*

We have presented a mathematical description of the mechanism of the PSOCT system in an earlier paper [17]. As is well-known, an arbitrary polarization state of light can be represented by the Stokes vector $\vec{S} \equiv [I \quad Q \quad U \quad V]^T$ where $I$ is the overall intensity, $Q$ denotes the intensity difference between vertical and horizontal linear polarizations, $U$ stands for the intensity difference between linear polarizations at +45° and -45°, and $V$ is the intensity difference between left and right circular polarizations. We showed in ref. 17 that the intensity of the output signal varies with the polarization state of the reference beam. The complete Stokes vector of backscattered object light can be expressed by linear combinations of the output signals, written as follows,

$$\vec{S}_r = [I'_H + I'_V, I'_H - I'_V, I'_P - I'_M, I'_R - I'_L]^T \qquad (1)$$

where $I'$ denotes the signal intensity after the LPF. The subscript stands for the polarization state of the reference beam, including vertical, horizontal, linear 45°, linear −45°, right circular and left circular, denoted as V, H, P, M, R and L, respectively.

The Mueller matrix, $M$, relates the Stokes vector of the incident object beam ($\vec{S}_o$) to the Stokes vector of the scattered light ($\vec{S}_r$): $\vec{S}_r = M\vec{S}_o$. Based on the acquisition of $\vec{S}_r$ according to eqn.(1), we can fully determine the 16 elements of $M$ by changing $\vec{S}_o$ via rotation of HWP2 and QWP2 in Fig.1. A detailed analysis shows that the full Mueller matrix is given by

$$M = \begin{bmatrix} I'_{HH}+I'_{HV}+I'_{VH}+I'_{VV} & I'_{HH}+I'_{HV}-I'_{VH}-I'_{VV} & I'_{PH}+I'_{PV}-I'_{MH}-I'_{MV} & I'_{RH}+I'_{RV}-I'_{LH}-I'_{LV} \\ I'_{HH}-I'_{HV}+I'_{VH}-I'_{VV} & I'_{HH}-I'_{HV}-I'_{VH}+I'_{VV} & I'_{PH}-I'_{PV}-I'_{MH}+I'_{MV} & I'_{RH}-I'_{RV}-I'_{LH}+I'_{LV} \\ I'_{HP}-I'_{HM}+I'_{VP}-I'_{VM} & I'_{HP}-I'_{HM}-I'_{VP}+I'_{VM} & I'_{PP}-I'_{PM}-I'_{MP}+I'_{MM} & I'_{RP}-I'_{RM}-I'_{LP}+I'_{LM} \\ I'_{HR}-I'_{HL}+I'_{VR}-I'_{VL} & I'_{HR}-I'_{HL}-I'_{VR}+I'_{VL} & I'_{PR}-I'_{PL}-I'_{MR}+I'_{ML} & I'_{RR}-I'_{RL}-I'_{LR}+I'_{LL} \end{bmatrix} \qquad (2)$$

where the first/second letter of the subscript denotes the polarization state of the incident object/reference beam, respectively.

**Stokesmetric imaging of the sample**

We applied the PSOCT system to perform Stokesmetric imaging of layers at varying depths in a porcine tendon sample. The imaged area (2.4mm ×2.0 mm) of the sample is chosen to consist of both muscle and fat. The incident object beam is set to be left-circularly polarized. The beam has a diameter of 40 microns after being focused by the lens. Two-dimensional image scanning is performed to three layers of different depths in the sample: (1) the surface, (2) the sub-surface layer at depth of 30 microns and (3) the sub-surface layer at a depth of 50 microns. The Stokes vector of backscattered light collected at each scanning step, treated as a pixel, is analyzed according to eqn.(1) and plotted in Fig.2. For *Q, U* and *V,* green and red colors stand for positive and negative values, respectively. The fifth column shows the unpolarized part of the light.

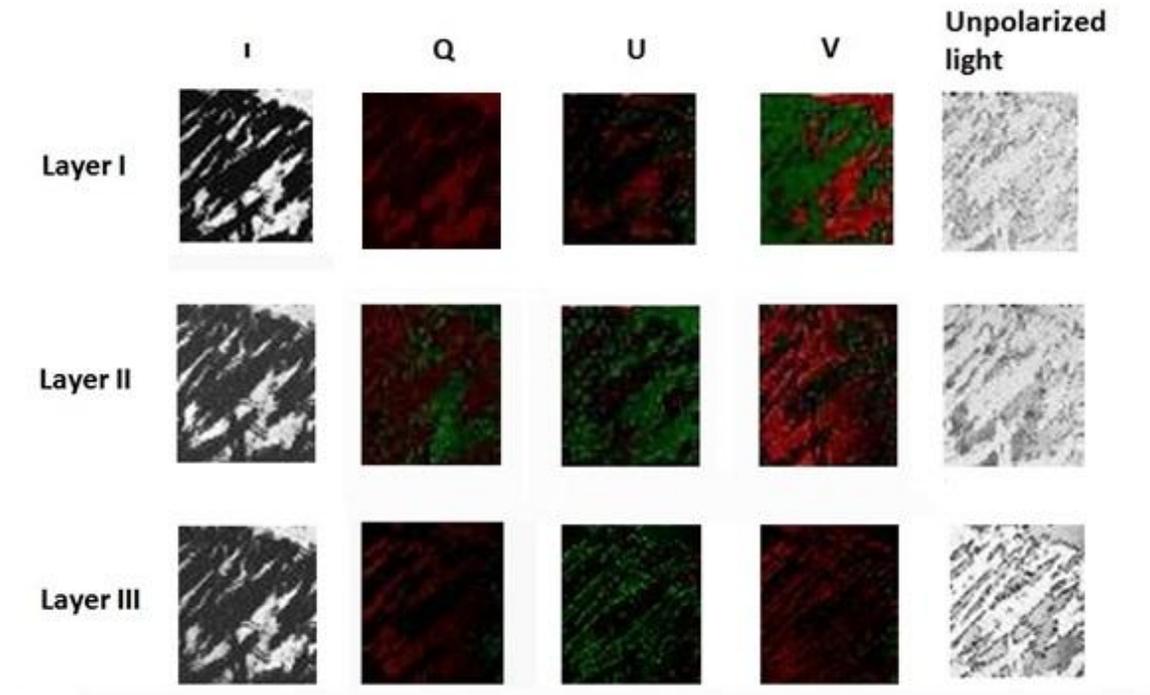

*Fig.2. Stokesmetric imaging of different layers within the porcine muscle sample.*

In Table-1, we present an analysis of this data for signals integrated over the whole image. Here we have defined the degree of depolarization as $DOD = (I_{unpol}/I) \times 100\%$. Note that these numbers very nearly satisfy the relation that $I_{unpol} = I - (|Q|^2 + |U|^2 + |V|^2)^{1/2}$. As can be seen from Table-1, the DOD rises as the depth of penetration in the sample increases. The table also shows that each layer couples the original polarization, which is purely circular, into linear polarizations to different extents. More importantly, as can be seen from Fig-2, the signs of the images for the Q and U elements, representing the directions of the linear polarizations, differ from one another for the three layers. As shown in ref.[19] and [20], the change of the polarization of light in a medium has a dependence on the geometrical and optical properties of the particles. The differences of signs and values of the reflected Q and U images of the three layers may imply different compositions of particles.

*Table-1 The DOD and proportion of polarization components of Stokes Vector for different layers.*

|  | DOD (%) | $|Q|^2/I^2$ (%) | $|U|^2/I^2$ (%) | $|V|^2/I^2$ (%) |
| --- | --- | --- | --- | --- |
| Layer I | 31.63 | 8.12 | 5.47 | 33.15 |
| Layer II | 44.90 | 13.77 | 7.60 | 8.98 |
| Layer III | 64.43 | 2.80 | 2.09 | 7.76 |

**Backscattering Muellermetric imaging**

We used the PSOCT to obtain the Muellermetric images of Layer III of the porcine tendon sample. In keeping with eqn. (2), thirty-six heterodyned images for different polarization combinations of the incident object beam and the reference beam are recorded. The sixteen images corresponding to the 4×4 backscattered Mueller matrix are computed by linear combinations of the raw images according to eqn.(2). These are shown below in Fig. 3 where $M_{ij}$ represents the *i*th row and the *j*th column of the Mueller matrix. Here again, green and red colors represent the positive and negative values, respectively.

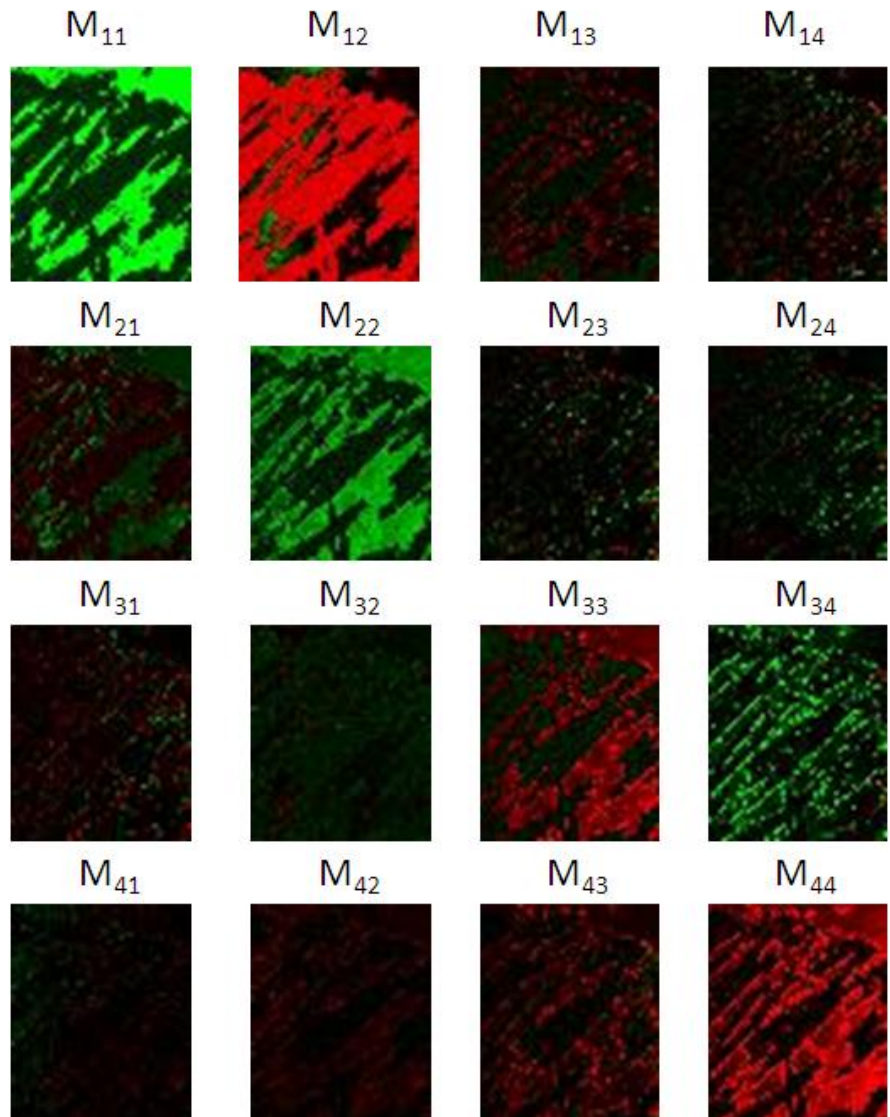

*Fig. 3. Backscattering Mueller matrix images of the porcine tendon sample*

Fig. 3 leads to some important observations about the backscattering Mueller matrix of this layer of the tendon sample. (1) The diagonal elements have much bigger magnitudes than the off-diagonal ones. Strong morphological patterns can be observed in the diagonal images. $M_{33}$ and $M_{44}$ exhibit negative values due to the π phase-shift produced at reflection. (2) For the off-diagonal elements, $M_{21}$, $M_{31}$ and $M_{41}$ have near-zero amplitudes, which agrees with the fact that the sample can hardly convert unpolarized light into polarized light. The elements $M_{34}$ and $M_{43}$, representing the coupling between circularly polarized light and 45° linearly polarized light, have significantly larger values than the other off-diagonal elements. The anti-symmetry (equal amplitude with opposite signs) of $M_{34}$ and $M_{43}$ indicate the birefringence produced by the sample. The phase delay between the horizontal and vertical polarizations turns the incident left-circularly polarized light into -45° linearly polarized light. In addition, Some dotted patterns are observed on $M_{34}$. The discontinuity may imply compositional variation on the layer.

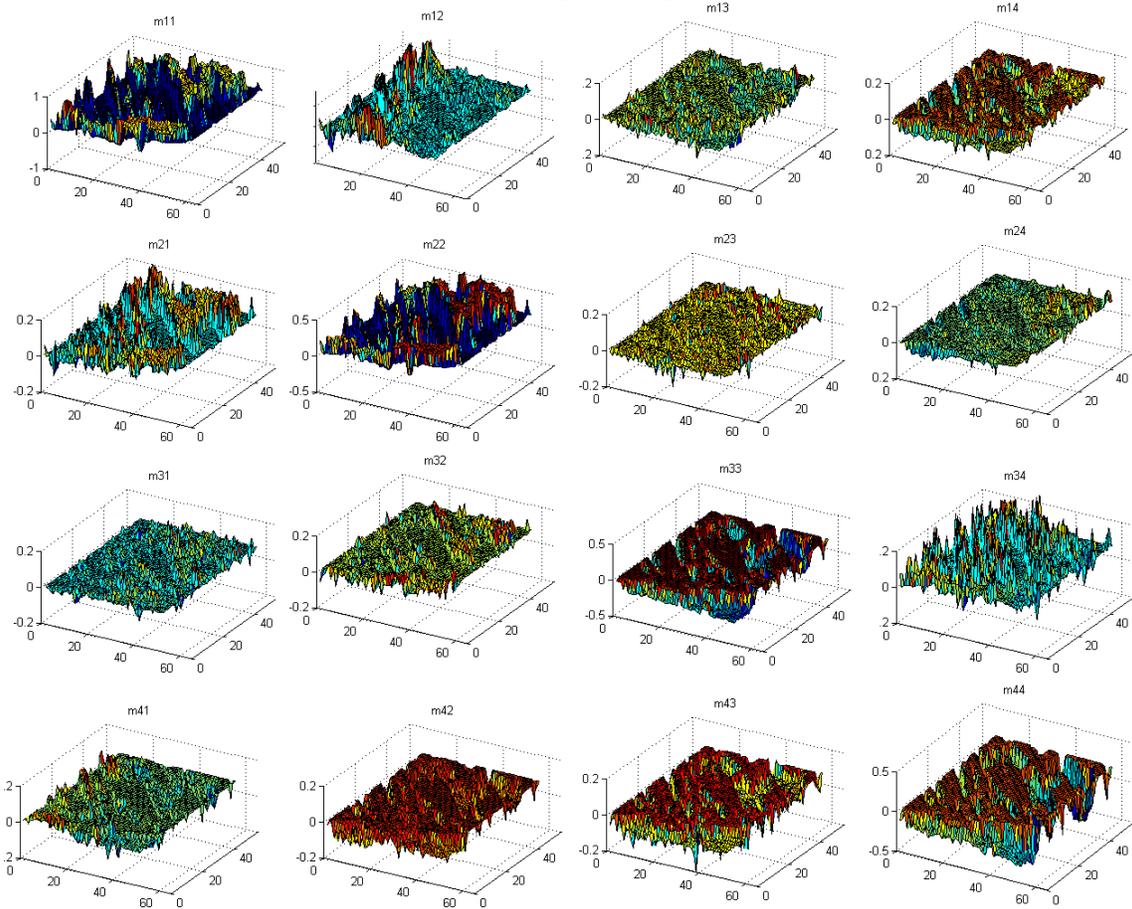

*Fig. 4. Axonometric plots of Mueller matrix images of the tendon sample for the third layer.*

The Muellermetric images of Layer III are plotted axonometrically in Fig. 4. The variations in spatial structures within each elementary image can be observed more easily in such a rendering.

**Conclusion**

To summarize, we built a PSOCT system based on heterodyning and filtering to perform, to the best of our knowledge, the first complete Muellermetric images of the porcine tendon sample. The effect of the different layers of the sample on scattering the incident light is discussed. The images reveal some information indiscernible from regular OCT. This study is very preliminary. A systematic study of a wide range of samples, coupled with theoretical modeling of the polarization sensitive scattering properties, is needed in order to establish the feasibility of using such an imaging system for applications such as clinical diagnosis of diseases.

**Acknowledgements**

This work is supported in part by AFOSR grant #FA9550-06-1-0466, NASA Grant #NNX09AU90A, and DOE grant #DE-AC02-06CH11357.